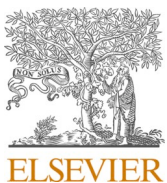
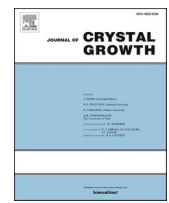

# Growth of millimeter-sized high-quality CuFeSe$_2$ single crystals by the molten salt method and study of their semiconducting behavior

Mingwei Ma [a,c,*], Binbin Ruan [a], Menghu Zhou [a], Yadong Gu [a], Qingxin Dong [a,b], Qingsong Yang [a,b], Qiaoyu Wang [a,d], Lewei Chen [a,b], Yunqing Shi [a,b], Junkun Yi [a,b], Genfu Chen [a,b], Zhian Ren [a,b]

[a] *Beijing National Laboratory for Condensed Matter Physics, Institute of Physics, Chinese Academy of Sciences, Beijing 100190, China*
[b] *University of Chinese Academy of Sciences, Beijing 100049, China*
[c] *Songshan Lake Materials Laboratory, Dongguan, Guangdong 100049, China*
[d] *Center for Advanced Quantum Studies and Department of Physics, Beijing Normal University, Beijing 100875, China*

A B S T R A C T

An eutectic AlCl$_3$/KCl molten salt method in a horizontal configuration was employed to grow millimeter-sized and composition homogeneous CuFeSe$_2$ single crystals due to the continuous growth process in a temperature gradient induced solution convection. The typical as-grown CuFeSe$_2$ single crystals in cubic forms are nearly 1.6 × 1.2 × 1.0 mm$^3$ in size. The chemical composition and homogeneity of the crystals was examined by both inductively coupled plasma atomic emission spectroscopy and energy dispersive spectrometer with Cu:Fe:Se = 0.96:1.00:1.99 consistent with the stoichiometric composition of CuFeSe$_2$. The magnetic measurements suggest a ferrimagnetic or weak ferromagnetic transition below $T_C$ = 146 K and the resistivity reveals a semiconducting behavior and an abrupt increase below $T_C$.

## 1. Introduction

Recently, CuFeSe$_2$ has drawn considerable attention due to its promising applications in thermoelectric technology [1–4], cancer phototherapy [5–8], energy storage systems [9], optoelectronics and solar energy conversion [10–13] as well as their natural abundance, low cost, environmentally friendly nature. It crystallizes in a tetragonal phase with space group $P\bar{4}2c$ and lattice parameters $a = b = 5.53$ Å and $c = 11.049$ Å, which is a superstructure based upon a cubic close-packed array of anions with the cations occupying a fraction of the available tetrahedral sites as shown in Fig. 1(a) [14]. There are still some controversies about the physical properties of CuFeSe$_2$. Bulk eskebornite CuFeSe$_2$ presents a metallic character and a magnetically ordered phase below $T_C$ = 70 K [15–18] whereas serveal reports on CuFeSe$_2$ film and nanocrystals reveal a semiconducting behavior [2,19]. Mössbauer absorption measurements indicate a ground state of spin-density waves (SDW) nature in CuFeSe$_2$ [20] whereas powder neutron scattering and hysteresis loop demonstrate a weak ferrimagnetic behavior [10,18,21,22]. A recent first-principle study reveals that CuFeSe$_2$ is a direct bandgap semiconductors with forbidden band widths of 0.64 eV [23] compared with a narrower bandgap of 0.45 eV in CuFeSe$_2$ nanoparticles [11]. The difference of the physical properties is probably ascribed to various sample defects or grain boundary effects. The single crystal of high quality is the prerequisite for studying the intrinsic physical properties of CuFeSe$_2$ single crystal which is not reported up to now.

There are several attempts on the crystal growth of CuFeSe$_2$. The phase transition in solid state of CuFeSe$_2$ occurs at 570 °C followed by a peritectic dissociation temperature 630 °C and a full melting point at 670 °C [24], implying the incongruent melting of tetragonal CuFeSe$_2$. Incongruent melting is a process occurring at temperature at which one solid phase transforms to another solid phase and a liquid phase with different chemical compositions than the original composition. CuFeSe$_2$ crystal was first grown by direct fusion of the elements in stoichiometric proportions or powder sample in a sealed evacuated quartz ampoule heated above its peritectic dissociation temperature [14,24], which will inevitably produce the dissociated phase inside CuFeSe$_2$ crystals, which is a higher temperature phase of CuFeSe$_2$ with spinel-type pseudocubic lattice phase [24]. The optimal temperature of crystal growth has to be below the phase transition temperature 570 °C such that a low temperature growth method with flux or transport agent is indispensible for the perfect single crystal. Single crystals of $CuFeSe_2$ grown by the gas transport reaction method with $I_2$ as a transporting agent are 0.7–0.8 cm in size accompanied by intermediate substances FeSeI and $Cu_3(SeI)_x$ with comparable crystal sizes ($0.8 \times 0.4 \times 0.005$ mm$^3$) [24]. However, they are crystallized in tetragonal lattice with larger parameters $a = b =$ 5.91 Å and $c = 11.82$ Å and space group of symmetry $I\bar{4}2d$ different from the space group $P\bar{4}2c$ of eskebornite $CuFeSe_2$. So, another key to a perfect crystal growth is the selection of flux or transport agent which does not react with the raw materials Fe, Cu or Se. $AlCl_3$/KCl molten salt is a good choice as its melting point is 120 °C and no other intermediate substances are produced when Fe, Cu or Se dissolved in AlCl$_3$/KCl solution. In this article we report an eutectic AlCl$_3$/KCl molten salt method to grow millimeter-sized and composition homogeneous CuFeSe$_2$ single crystals. The typical as-grown CuFeSe$_2$ single crystal was nearly 1.6 × 1.2 × 1.0 mm$^3$ in size. In contrast to the previous magnetic and resistivity studies [15–18], our CuFeSe$_2$ single crystals reveal a semiconducting behavior and a ferrimagnetic or weak ferromagnetic transition below $T_C = 146$ K.

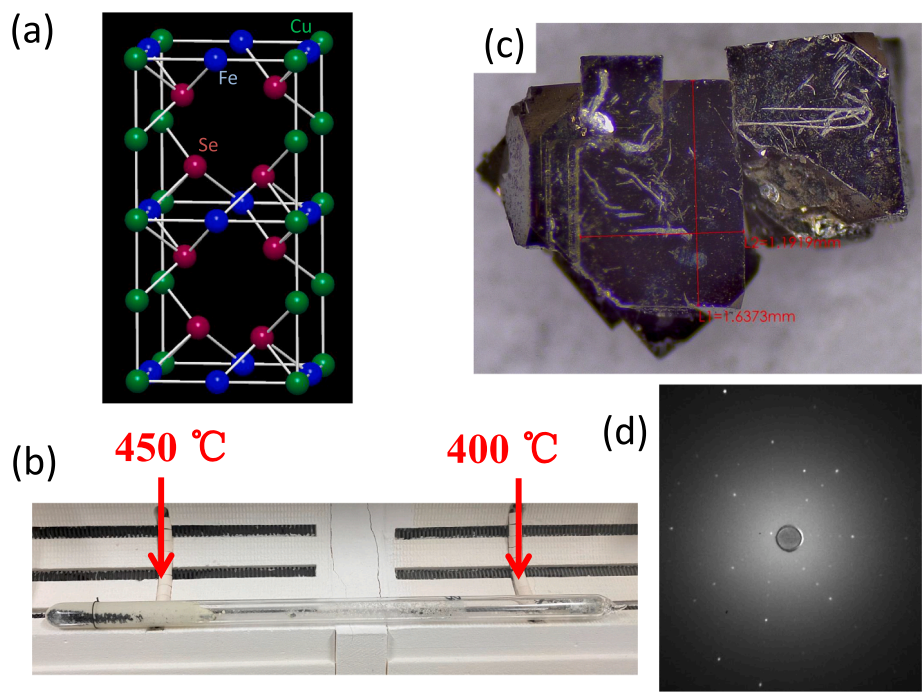

**Fig. 1.** (a) Crystal structure of $CuFeSe_2$. (b) Experimental schemes of the horizontal configuration for a convective $AlCl_3$/KCl molten salt method by which millimeter-sized $CuFeSe_2$ single crystals have been grown. (c) Photograph of typical $CuFeSe_2$ single crystals with cubic shapes and dimensions about $1.6 \times 1.2 \times 1.0$ mm$^3$ grown in present work. (d) Laue picture of $CuFeSe_2$ single crystal.

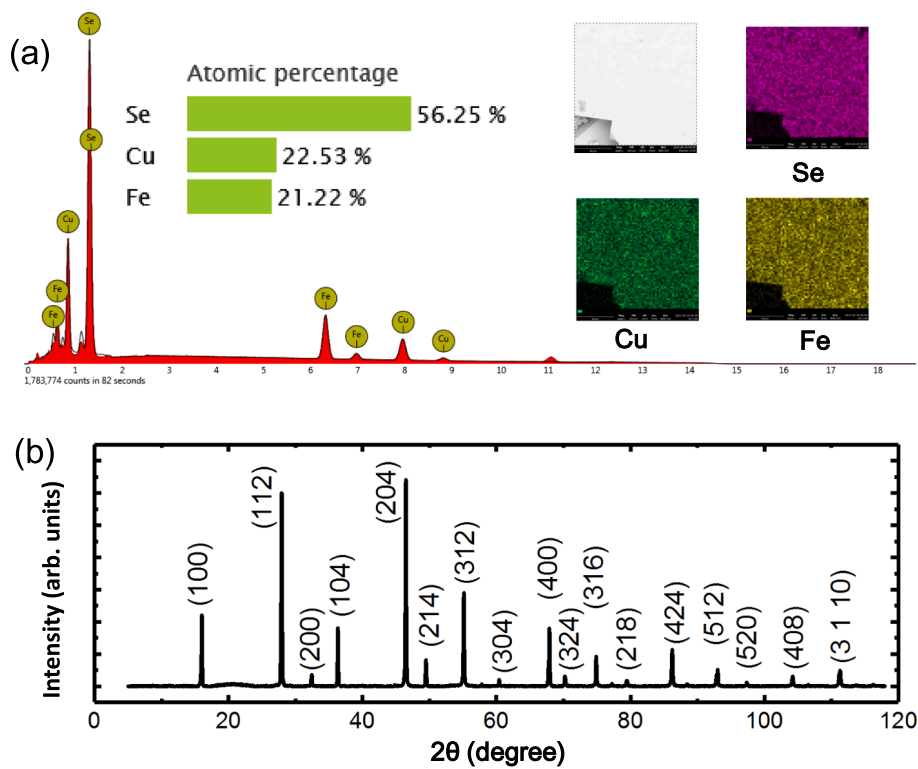

**Fig. 2.** (a) The EDS microanalysis spectrum taken on the surface area of $CuFeSe_2$ single crystal. The inset shows the SEM image taken from the flat surface of $CuFeSe_2$ single crystal and elements Se, Fe, Cu distribution over the surface of $CuFeSe_2$ single crystal. (b) Powder x-ray diffraction patterns of $CuFeSe_2$.

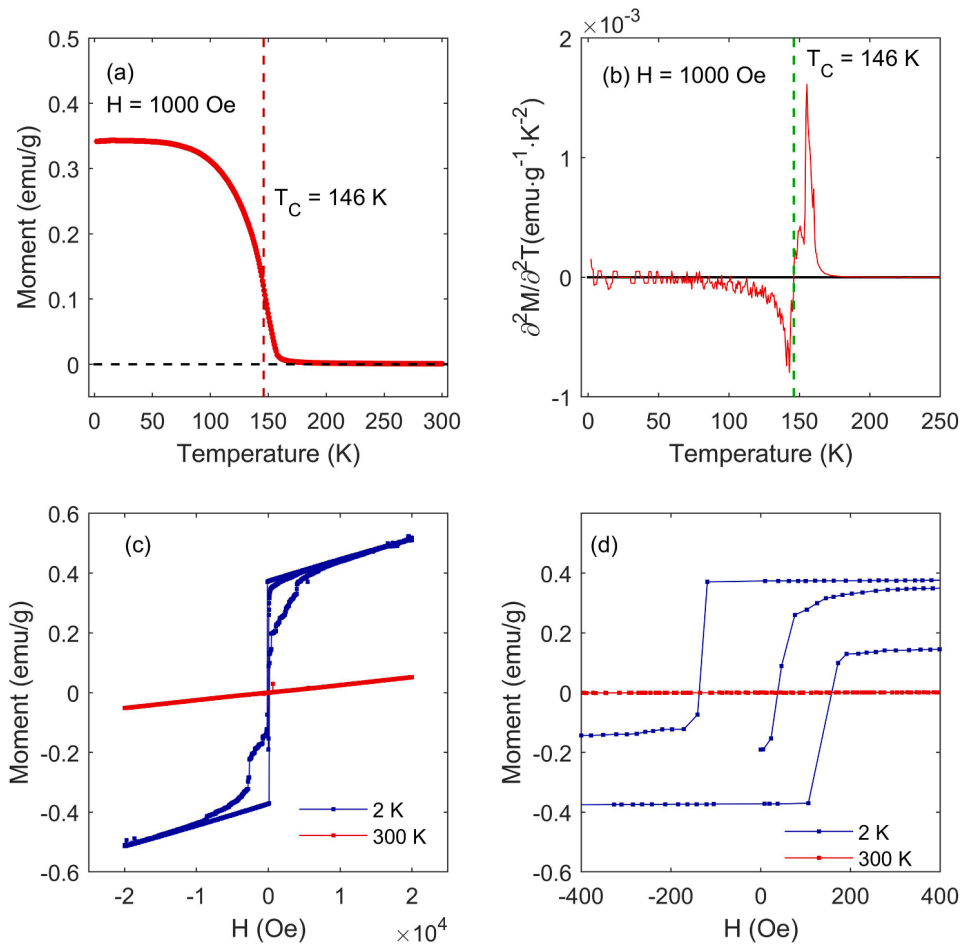

**Fig. 3.** (a) Temperature dependence of magnetization of CuFeSe$_2$. (b) The derivative of magnetization $\partial^2 M/\partial^2 T$ versus temperature. (c) Hysteresis loop at 2 K and 300 K. (d) An enlarged view of the hysteresis loop around $H = \pm 400$ Oe.

## 2. Experimental methods

The raw materials of Fe, Se and Cu powder with molar ratio Fe:Cu:Se = 1:1:2 are mixed together with the salts AlCl$_3$:KCl = 2:1 in molar ratio and sealed in an evacuated quartz tube ($\Phi$10 mm × 300 mm). The quartz tube was placed in a horizontal double zone furnace with temperature 450 °C and 400 °C, respectively, resulting in a stable temperature gradient 2 °C/cm as displayed in Fig. 1(b). After a growth duration of 30 days, a large number of high-quality and composition homogeneous CuFeSe$_2$ single crystals with cubic forms appeared around the cold part of quartz tube. The CuFeSe$_2$ single crystals are extracted by dissolving the AlCl$_3$/KCl solvent in distilled water. The typical crystal sizes are up to 1.6 × 1.2 × 1.0 mm$^3$ with regular and flat surfaces as shown in Fig. 1 (c). Fig. 1(d) shows an example of back-scattering x-ray Laue patterns obtained on (0 0 1) plane (*ab* plane) of CuFeSe$_2$ single crystal. The result shows very sharp and uniform diffraction spots with the 4-fold axis symmetry demonstrating the crystalline perfection of the CuFeSe$_2$ single crystal without twin crystals.

It should be emphasized that conventional flux growth is usually carried out in a vertical configuration of the furnace. The single crystals are grown by slow cooling the supersaturated solution. In our case, CuFeSe$_2$ has low solubility in AlCl$_3$/KCl solution due to the limitation of the lower growth temperature. Also, the narrower temperature range for crystallization is the obstacle to a continuous growth process of CuFeSe$_2$ single crystals. The horizontal configuration with stable temperature gradient can effectively overcome or compensate the above limitations at the expense of longer growth duration of 30 days.

The micro-morphology of CuFeSe$_2$ single crystals was examined by scanning electron microscope (SEM) on Phenom ProX. The chemical composition was determined by both inductively coupled plasma atomic emission spectroscopy (ICP) and energy dispersive spectrometer (EDS) on Phenom ProX. Powder x-ray diffraction measurements were carried out at room temperature on an x-ray diffractometer (Rigaku UltimaIV) using Cu K$_\alpha$ radiation. The resistance of crystal sample was measured on Quantum Design PPMS-9 using the standard 4-probe method. The magnetic measurements were carried out on a SQUID magnetometer (Quantum Design MPMS XL-1).

## 3. Results and discussion

The chemical composition and homogeneity of the crystals was examined by the EDS analysis and one of the typical EDS spectrums taken on the crystal surface is shown in Fig. 2(a). The composition of the crystal is very homogeneous as demonstrated in the insets of Fig. 2(a). The mapping areas with purple, green and yellow represent the homogeneous distribution of Se, Cu and Fe, respectively. The chemical composition is roughly estimated to be Cu:Fe:Se = 22.53:21.22:56.25 ≈ 0.90:0.85:2.25. The deviation of stoichiometry of Cu:Fe:Se = 1:1:2 is probably attributed to the measurement only on the surface of crystals. For a more accurate determination of the chemical composition, several crystals are ground into powder measured by ICP, which gives the molar ratio with Cu:Fe:Se = 0.96:1.00:1.99, which is consistent with the stoichiometric composition of $CuFeSe_2$. Fig. 2(b) shows the powder x-ray diffraction (XRD) patterns of $CuFeSe_2$ at room temperature. Powder XRD demonstrates that all the diffraction peaks of $CuFeSe_2$ can be well indexed with a previously reported tetragonal structure with the space group of $P\bar{4}2c$ and lattice parameters $a = 5.5190(3)$ Å and $c = 11.0488(7)$ Å. No impurity peaks can be observed in our XRD spectrum indicating the high purity and quality of our $CuFeSe_2$ single crystals.

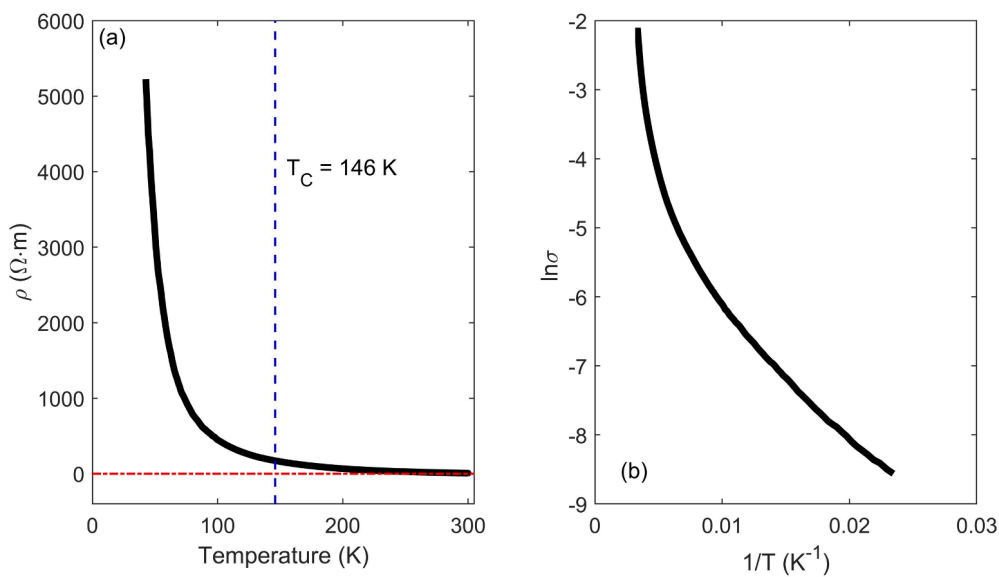

**Fig. 4.** (a) Temperature dependence of resistivity of $CuFeSe_2$. (b) Temperature dependence of conductivity ($\ln\sigma$ vs $1/T$) of $CuFeSe_2$.

As we know, the grain boundary in powder sample and impurity intergrown with $CuFeSe_2$ will conceal its intrinsic properties. After characterization of the high purity and quality of $CuFeSe_2$ single crystal, we turn to its magnetic and electrical properties. Fig. 3(a) shows the temperature dependence of magnetic susceptibility ranging from 2 K to 300 K which displays a paramagnetic and ferrimagnetic behaviour above and below $T_C$ respectively. Fig. 3(b) shows the derivative of magnetization $\partial^2 M/\partial^2 T$ versus temperature which suggest the $T_C \approx 146$ K. Previous reports on $CuFeSe_2$ sample showed paramagnetic behavior from room temperature down to $T_C \approx 70$ K or 80 K, below which the Fe atoms on the $2e$ and $2a$ sites (of the tetragonal cell) have slightly different magnetic moments, resulting in a weak ferrimagnetic behavior [15–18]. To verify the magnetic response of the $CuFeSe_2$ single crystals, the hysteresis loops were measured at 2 and 300 K and between −20,000 Oe and +20,000 Oe and the results are presented in Fig. 3(c). The hysteresis loop suggests a ferrimagnetic behaviour at 2 K and a paramagnetic behaviour at 300 K. At both temperatures, the hysteresis loop has a slanted appearance which is not saturated at 20,000 Oe, consistent with previous measurements on $CuFeSe_2$ nanocrystals [10,25]. The enlarged view between $H = \pm 400$ Oe is displayed in Fig. 3 (d), where the coercive force $H_c \approx 130$ Oe at $T = 2$ K in contrast to 1360 Oe or 1780 Oe on $CuFeSe_2$ nanocrystals [10,22]. Also, there still exist non-linear $S$-shape curve in hysteresis loop at room temperature in previous report indicating possible ferromagnetic impurity in their samples [10,22,25]. The linear behavior in our hysteresis loop at 300 K suggests the high purity of our as-grown $CuFeSe_2$ single crystals. Fig. 4 (a-b) shows the temperature dependence of resistivity and conductivity which reveals a semiconducting behavior in agreement with the results on $CuFeSe_2$ thin films or nanocrystals [2,19] instead of a metalic character in $CuFeSe_2$ bulk sample [15]. The differences in magnetic and electrical properties on various $CuFeSe_2$ samples synthesized via different preparation process is probably due to size effects, surface effects, interparticle interactions or impurities inside the previous sample.

## 4. Conclusion

We have grown millimeter-sized $CuFeSe_2$ single crystals using eutectic $AlCl_3/KCl$ molten salt in a quartz ampoule placed in a double temperature zone furnace of horizontal configuration. The clear Laue spots and linear behavior of the hysteresis loops at 300 K indicates the high quality of $CuFeSe_2$ single crystals without any other magnetic impurities. A ferrimagnetic or weak ferromagnetic transition was observed below $T_C = 146$ K where the resistivity shows an abrupt enhancement which reflect the intrinsic properties of $CuFeSe_2$ single crystals.

## CRediT authorship contribution statement

**Mingwei Ma:** Conceptualization, Methodology, Writing – review & editing. **Binbin Ruan:** Software. **Menghu Zhou:** Software. **Yadong Gu:** Software. **Qingxin Dong:** Software. **Qingsong Yang:** . **Qiaoyu Wang:** Software. **Lewei Chen:** . **Yunqing Shi:** . **Junkun Yi:** . **Genfu Chen:** Conceptualization, Methodology. **Zhian Ren:** Conceptualization, Methodology.

## Declaration of Competing Interest

The authors declare that they have no known competing financial interests or personal relationships that could have appeared to influence the work reported in this paper.

## Data availability

Data will be made available on request.

## Acknowledgments

We gratefully thank Lihong Yang, Hongjun Shi, Li Wang and Youting Song for their technical help in x-ray diffraction and chemical analysis. The work was supported by the National Natural Science Foundation of China (Grant No. 12004418), the National Key Research and Development of China (Grant No. 2018YFA0704200, 2022YFA1602800) and the Strategic Priority Research Program of Chinese Academy of Sciences

(Grant No. XDB25000000).